\documentclass[doublecol]{epl2} 

\usepackage{amsmath,amssymb} 
\usepackage{graphics,color}
\usepackage{subfig}

\title{The frustrated Heisenberg antiferromagnet on the honeycomb lattice: $J_{1}$--$J_{2}$ model}
\shorttitle{The $J_{1}$--$J_{2}$ honeycomb lattice} 

\author{P.~H.~Y.~Li\inst{1} 
\and R.~F.~Bishop\inst{1}
\and D.~J.~J.~Farnell\inst{2}   
\and C.~E.~Campbell\inst{3}}
\shortauthor{P.~H.~Y.~Li \etal}

\institute{               
  \inst{1} School of Physics \& Astronomy, Schuster Building, The University of Manchester, Manchester M13 9PL, UK\\   
  \inst{2} Division of Mathematics, Faculty of Advanced Technology, University of Glamorgan, Pontypridd CF37 1DL, UK \\  
  \inst{3} School of Physics \& Astronomy, University of Minnesota, 116 Church St. SE, Minneapolis, MN 55455, USA}


\abstract{
We study the ground-state (gs) phase diagram of the frustrated
  spin-$\frac{1}{2}$ $J_{1}$--$J_{2}$ antiferromagnet with
  $J_{2}=\kappa J_1>0$ ($J_{1}>0$) on the honeycomb lattice,
  using the coupled-cluster method.
  We present results for the ground-state energy, magnetic order
  parameter and plaquette valence-bond crystal (PVBC) susceptibility.  
  We find a paramagnetic PVBC phase for $\kappa_{c_1}<\kappa<\kappa_{c_2}$, where
  $\kappa_{c_1} \approx 0.207 \pm 0.003$ and $\kappa_{c_2} \approx 0.385 \pm 0.010$.
  The transition at $\kappa_{c_1}$ to the N\'{e}el phase seems to
  be a continuous deconfined transition (although we cannot exclude
  a very narrow intermediate phase in the range 
  $0.21 \lesssim \kappa \lesssim 0.24$), while that at
  $\kappa_{c_2}$ is of first-order type to another quasiclassical
  antiferromagnetic phase that occurs in the classical version of the model 
  only at the isolated and highly degenerate critical point $\kappa = \frac{1}{2}$.  
  The spiral phases that are present classically for all values
  $\kappa > \frac{1}{6}$ are absent for all $\kappa \lesssim 1$.
}
                      
\begin{document}

\maketitle

Two-dimensional (2D) frustrated quantum spin-lattice systems have become 
of huge interest both theoretically and experimentally
\cite{Senthil:2004,Moessner:2006,SRFB:2004}.  Attention has particularly focussed 
on the rich panoply of (zero-temperature, $T=0$) quantum phase transitions 
that they exhibit \cite{SRFB:2004,Sachdev_book:1999}.
Without thermal fluctuations the transitions are driven
solely by the interplay of quantum fluctuations and any frustration due to 
inherent competition between the interactions.  Such
frustration can arise either dynamically or geometrically.  A prototypical
example of the former is the well-studied $J_{1}$--$J_{2}$ Heisenberg
antiferromagnet (HAFM) on the bipartite square lattice (see, e.g., Refs. 
\cite{j1j2_square_ccm1,Bi:2008_PRB,j1j2_square_ccm5} and
references cited therein), where 
nearest-neighbour (NN) spins interact via a Heisenberg interaction with strength
parameter $J_{1}>0$, which competes with a Heisenberg interaction with 
strength parameter $J_{2}>0$ between next-nearest neighbour (NNN) pairs. 
Similar prototypical models exhibiting geometrical frustration are the 
pure NN HAFMs on the triangular \cite{Bernu:1994} and kagome
lattices \cite{Kagome_Schn:2008}.  
For either form of frustration special interest then centres on 
the possible appearance of novel quantum ground-state (gs) phases  
without the long-range order (LRO) that typifies the classical gs phases 
of the corresponding models taken in the limit $s \rightarrow \infty$ of 
the spin quantum number $s$ of the lattice spins.  Examples include
various valence-bond crystalline solid phases and 
spin-liquid phases.

Quantum fluctuations tend to be largest for the smallest values of $s$, for lower
dimensionality $D$ of the lattice, and for the smallest coordination of the lattice.
Thus, for spin-$\frac{1}{2}$ models in $D=2$, the honeycomb 
lattice plays a special role.  Frustration is easily incorporated via competing NNN and 
maybe also next-next-nearest-neighbour (NNNN) bonds.  Such models and their 
experimental realisations have been much studied in recent years 
\cite{honey1,honey2,exp,honey5,Cabra,honey7,Albuquerque:2011}.  
Additional interest has also sprung from the recent synthesis of graphene 
monolayers and other magnetic materials with a
honeycomb structure.  Theoretical interest was spurred by 
the discovery of a spin-liquid phase in the exactly solvable Kitaev model 
\cite{kitaev}, in which spin-$\frac{1}{2}$ particles reside on a honeycomb 
lattice.  Hubbard models on the honeycomb lattice may also describe many of the
relevant physical properties of graphene.  For example, evidence has been 
found \cite{meng} that quantum fluctuations are sufficiently 
strong to establish an insulating spin-liquid phase between the nonmagnetic 
metallic phase and the antiferromagnetic (AFM) Mott insulator phase, when the 
Coulomb repulsion parameter $U$ becomes moderately strong.  For large values of 
$U$ the latter phase corresponds to the pure HAFM on the bipartite honeycomb lattice, 
whose gs phase exhibits N\'{e}el LRO.  However, higher-order terms in the $t/U$
expansion of the Hubbard model may lead to frustrating exchange couplings 
in the corresponding spin-lattice limiting model, in which the HAFM with NN exchange
couplings is the leading term in the large-$U$ expansion.  There is a 
growing consensus \cite{honey1,honey2,honey5,Cabra,honey7,Albuquerque:2011} 
that the frustrated spin-$\frac{1}{2}$ HAFM on the honeycomb lattice 
undergoes a frustration-induced quantum phase
transition to a paramagnetic phase showing no magnetic LRO.

Indirect experimental backup for such theoretical findings comes from recent
observations of spin-liquid-like behaviour in the layered compound 
Bi$_{3}$Mn$_{4}$O$_{12}$(NO$_{3})$ (BMNO) at temperatures below its
Curie-Weiss temperature \cite{exp}.  In BMNO the Mn$^{4+}$ ions reside on
the sites of (weakly-coupled) honeycomb lattices, but they 
have spin quantum number $s=\frac{3}{2}$.  The successful replacement 
of the Mn$^{4+}$ ions in BMNO by V$^{4+}$ ions could lead to the
realization of a corresponding $s=\frac{1}{2}$ model on the honeycomb lattice. 
Other recent realizations of HAFMs on a honeycomb lattice include compounds such as
Na$_{3}$Cu$_{2}$SbO$_{6}$ \cite{Miura:2006} and InCu$_{2/3}$V$_{1/3}$O$_{3}$ 
\cite{Kataev:2005}, in both of which the Cu$^{2+}$ ions in the copper oxide layers form a 
spin-$\frac{1}{2}$ HAFM on a (distorted) honeycomb lattice.  Others
include the family of compounds BaM$_{2}$(XO$_{4}$)$_{2}$ (M$=$Co, Ni;
X=P, As) \cite{Regnault}, in which the magnetic ions M are disposed in
weakly-coupled layers where they reside on the sites of a honeycomb lattice. 
The Co ions have spins $s=\frac{1}{2}$ and the Ni ions have spins $s=1$.  Finally,
recent calculations of the low-dimensional material $\beta$-Cu$_{2}$V$_{2}$O$_{7}$
\cite{Tsirlin:2010} show that its magnetic properties can be described in terms of
a spin-$\frac{1}{2}$ model on a distorted honeycomb lattice.

\begin{figure}[!tb]
\mbox{
\subfloat[N\'{e}el]{\scalebox{0.25}{\includegraphics{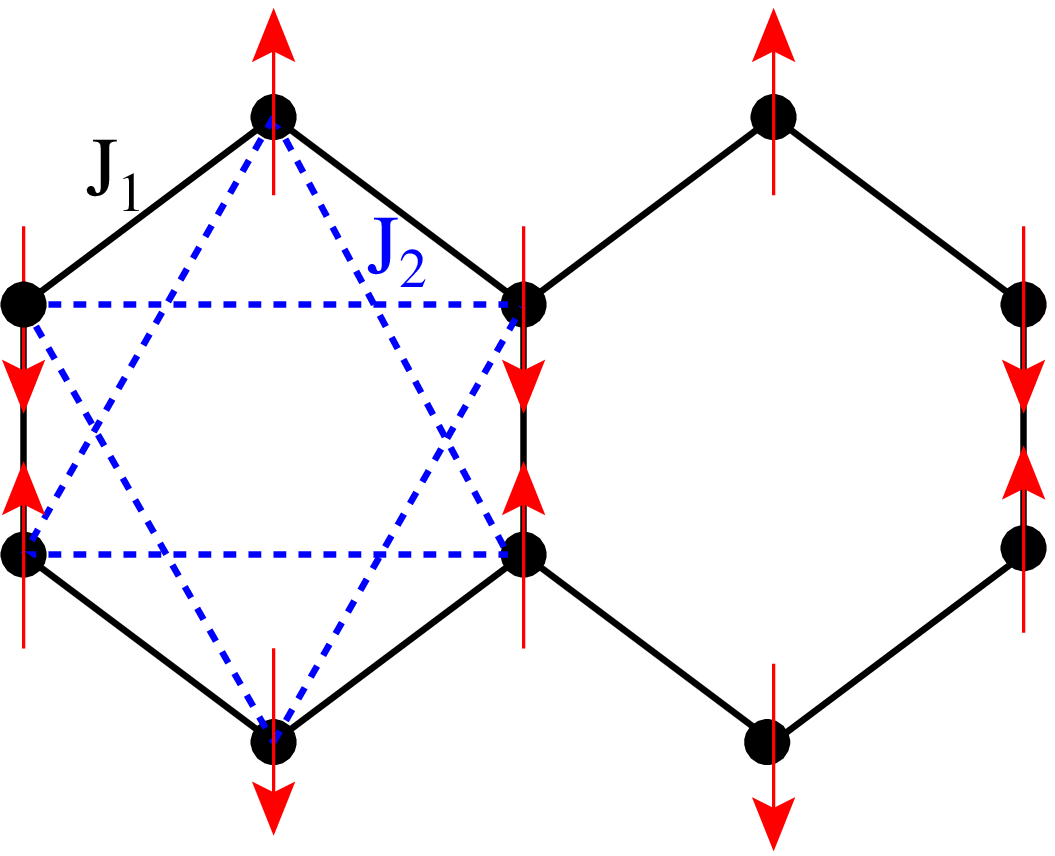}}}
\subfloat[spiral]{\scalebox{0.25}{\includegraphics{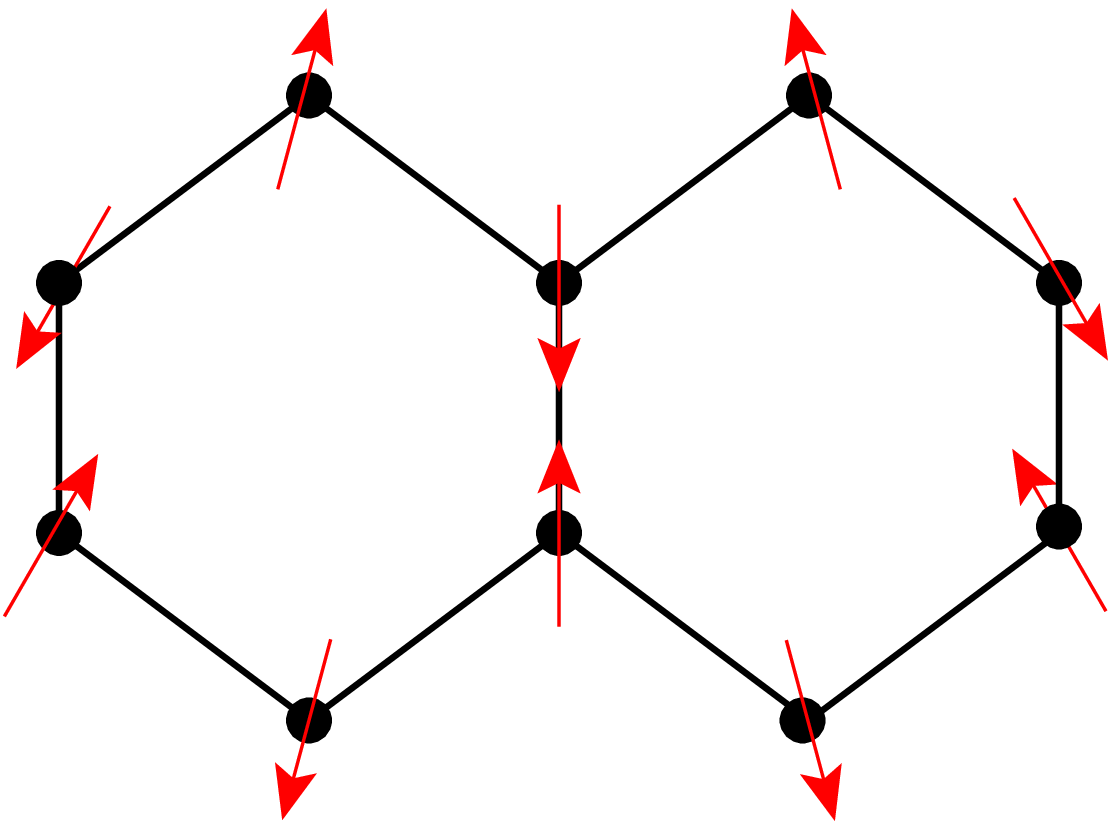}}}
\subfloat[anti-N\'{e}el]{\scalebox{0.25}{\includegraphics{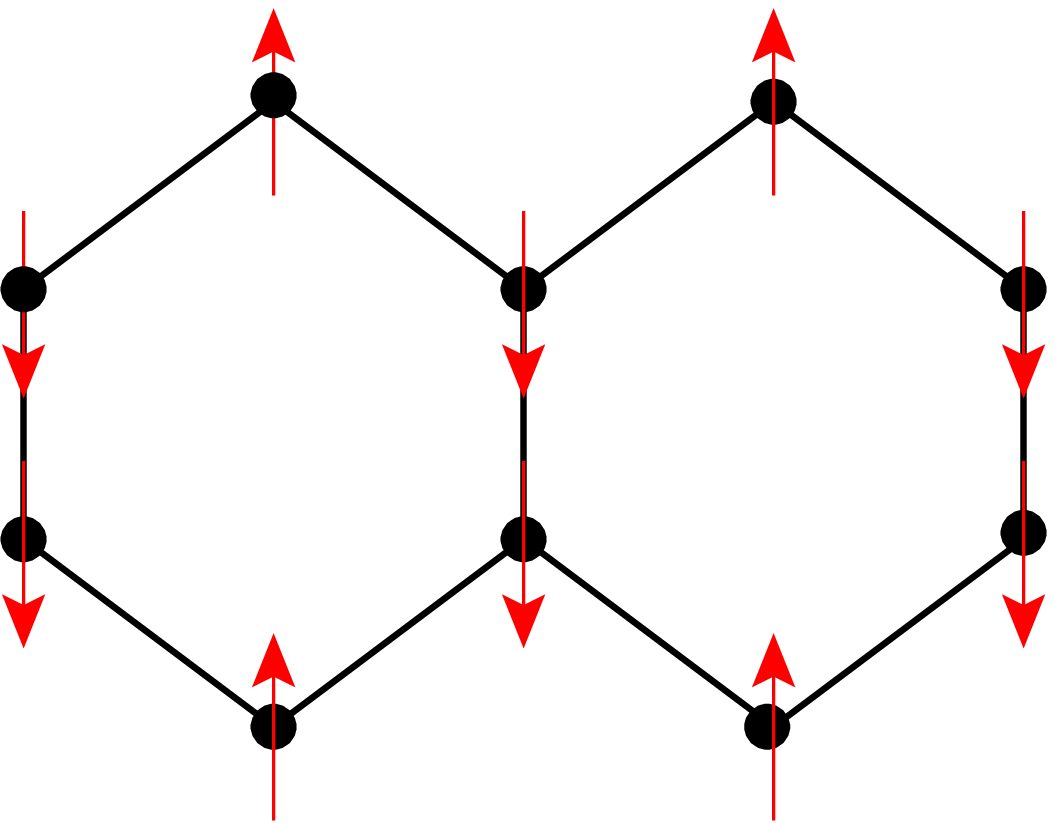}}}
}
\caption{(Color online) The $J_{1}$--$J_{2}$ model on the honeycomb lattice (with $J_{1}=1$),
  showing (a) the N\'{e}el, (b) spiral, and (c) anti-N\'{e}el states. 
  The arrows represent spins located on lattice sites \textbullet.}
\label{model}
\end{figure}

In recent papers \cite{DJJF:2011_honeycomb,PL:2011_Honeycomb_FM} 
we have studied the frustrated spin-$\frac{1}{2}$ 
$J_{1}$--$J_{2}$--$J_{3}$ model on the honeycomb lattice 
\cite{honey1,honey2,honey5,Cabra,honey7,Albuquerque:2011}.  
Its Hamiltonian is given by
\begin{equation}
H = J_{1}\sum_{\langle i,j \rangle} \mathbf{s}_{i}\cdot\mathbf{s}_{j} + J_{2}\sum_{\langle\langle i,k \rangle\rangle} 
\mathbf{s}_{i}\cdot\mathbf{s}_{k} + J_{3}\sum_{\langle\langle\langle i,l \rangle\rangle\rangle} \mathbf{s}_{i}\cdot\mathbf{s}_{l}\,,
\label{eq1}
\end{equation}
where index $i$ runs over all honeycomb lattice sites, and indices $j$, $k$, and $l$ run over all
NN, NNN and NNNN sites to $i$, respectively, counting each bond once
and once only.  Each lattice site $i$ carries a particle with spin 
$s=\frac{1}{2}$ and a spin operator ${\bf s}_{i}=(s_{i}^{x},s_{i}^{y},s_{i}^{z})$. 
The lattice and exchange bonds are illustrated in Fig.~\ref{model}.
In earlier work we restricted ourselves to the case where
$J_{3}=J_{2}>0$, but we dealt both with the AFM case with
$J_{1}>0$ \cite{DJJF:2011_honeycomb} and the ferromagnetic (FM) case with
$J_{1}<0$ \cite{PL:2011_Honeycomb_FM}.  Here we put $J_{3}=0$, 
and hence consider the $J_{1}$--$J_{2}$ model.  We restrict ourselves to
the case where both bonds are of AFM type, $J_{1}>0$ and
$J_{2} \equiv \kappa J_{1} > 0$.  We henceforth set $J_{1} \equiv 1$.

The classical ($s \rightarrow \infty$) gs phase diagram of the $J_{1}$--$J_{2}$--$J_{3}$ model 
on the honeycomb lattice \cite{Rastelli:1979,honey2} comprises six different
phases when $J_{1}>0$ and the other two bonds, $J_2$ and $J_3$, can take either sign.  
Three are collinear AFM phases, one is the FM phase, and the 
other two are different helical phases (and see, e.g., Fig. 2 of Ref. \cite{honey2}). 
The AFM phases are the N\'{e}el phase (N) shown in Fig.~\ref{model}(a), the striped (S) phase discussed
in our earlier paper \cite{DJJF:2011_honeycomb}, and the anti-N\'{e}el (aN) phase shown
in Fig.~\ref{model}(c).  The S, aN, and N states have, respectively, 1, 2, and all 3 NN spins
to a given spin antiparallel to it.  Equivalently, if we consider the sites of the honeycomb lattice
as comprising a set of parallel sawtooth (or zigzag) chains (in any one of the three equivalent
directions), the S state comprises alternating FM chains, while the aN state comprises AFM chains 
in which NN spins on adjacent chains are parallel.  Although there are infinite
manifolds of non-coplanar states degenerate in energy with each of the S and aN states at $T=0$, 
both thermal and quantum fluctuations \cite{honey2} select the collinear configurations.  
When $J_{3}>0$ there is a region in which the spiral state shown in Fig.~\ref{model}(b) is the
stable gs phase.  It is characterized by a spiral angle defined so that as we move along
the parallel sawtooth chains [drawn in the horizontal direction in Fig.~\ref{model}(b)]
the spin angle increases by $\pi + \phi$ from one site to the next, 
and with NN spins on adjacent chains antiparallel.  The classical gs energy is minimized 
for this spiral state when the pitch angle $\phi = \cos ^{-1} [\frac{1}{4}(J_{1}-2J_{2})/(J_{2}-J_{3})]$,
when the energy per spin takes the value,
\begin{equation}
\frac{E^{\rm cl}_{\rm spiral}}{N} = \frac{s^{2}}{2}\left (-J_{1} - 2J_{2} + J_{3} - \frac{1}{4} 
\frac{(J_{1} - 2J_{2})^{2}}{(J_{2} - J_{3})} \right ). 
\end{equation}

We note that as $\phi \rightarrow 0$ this spiral state becomes the collinear N state with
energy per spin, 
\begin{equation}
\frac{E^{\rm cl}_{\rm N}}{N} = \frac{s^{2}}{2}\left (-3J_{1} + 6J_{2} - 3J_{3}\right), 
\end{equation}
and there is a continuous phase transition between these two states on the boundary
$y=\frac{3}{2}x-\frac{1}{4}$, for $\frac{1}{6}<x<\frac{1}{2}$, where $y\equiv J_{3}/J_{1}$ 
and $x\equiv J_{2}/J_{1}$.  Similarly, as $\phi \rightarrow \pi$ the spiral state becomes 
the collinear S state, and there is a continuous phase transition between the 
two states on the boundary line $y=\frac{1}{2}x+\frac{1}{4}$, for
$x>\frac{1}{2}$.  There is a first-order phase transition between the collinear N and
S states along the boundary line $x=\frac{1}{2}$, for $y>\frac{1}{2}$.  These three
phases (N, S, and spiral) meet at the tricritical point $(x,y) = (\frac{1}{2},\frac{1}{2})$.
As $x \rightarrow \infty$ (for fixed finite $y$), the 
spiral pitch angle $\phi \rightarrow \frac{2}{3} \pi$.  In this limit the model becomes
two HAFMs on weakly connected interpenetrating triangular lattices, with the usual 
classical ordering of NN spins oriented at an angle $\frac{2}{3} \pi$ to each
other on each sublattice.

The above three states are the only classical gs phases when $y>0$.  For $y<0$ the N state
persists in a region bounded by the same boundary line as above, 
$y=\frac{3}{2}x-\frac{1}{4}$, for $-\frac{1}{2}<x<\frac{1}{6}$, on which it continuously
meets a second spiral state, and by the boundary line $y=-1$, for $x<-\frac{1}{2}$, at which 
it undergoes a first-order transition to the FM state, which itself is the stable gs phase
in the region $x<-\frac{1}{2}$ and $y<-1$.  Another collinear AFM state, the 
aN state shown in Fig.~\ref{model}(c), with energy per spin,
\begin{equation}
\frac{E^{\rm cl}_{\rm aN}}{N} = \frac{s^{2}}{2}\left (-J_{1} - 2J_{2} + 3J_{3}\right), 
\end{equation}
becomes the stable gs phase in the region $x>\frac{1}{2}$, for 
$y < \frac {1}{2}\{x -[x^{2}+2(x-\frac{1}{2})^{2}]^{1/2} \}$.  On the boundary it
undergoes a first-order transition to the spiral state shown in Fig.~\ref{model}(b).
For $\frac{1}{6}<x<\frac{1}{2}$ the spiral state shown in Fig.~\ref{model}(b) meets
another spiral gs phase on the boundary line $y=0$, at which point there is a 
first-order transition.  The pitch angle of this second spiral phase smoothly
approaches the value zero along the above boundary with the N state, and the 
value $\pi$ along a second boundary curve that joins the points $(x,y) = (-\frac{1}{2},-1)$ 
and $(\frac{1}{2},0)$, on which it meets the aN state.  Both transitions are continuous ones.
This second spiral meets the three collinear states N, aN, and FM at the tetracritical
point $(x,y) = (-\frac{1}{2},-1)$.

Henceforth we restrict consideration to the $J_{1}$--$J_{2}$ model where $J_{3}=0$ (and 
$J_{1} \equiv 1$).  The classical ($s \rightarrow \infty$) model thus has the N state as 
its gs for $J_{2} < \frac{1}{6}$, whereas for $J_{2} > \frac{1}{6}$ the gs comprises
an infinite family of degenerate coplanar states with spiral order 
[including that shown in Fig.~\ref{model}(b)], in which the 
spiral wave vector can point in any direction \cite{Rastelli:1979,honey2,Mulder:2010}. 
It is found \cite{Mulder:2010} that, to leading order in $1/s$, spin wave fluctuations
lift this degeneracy by the well-known order-by-disorder mechanism, in favour of specific
wave vectors.  However, these spiral states for the $J_{1}$--$J_{2}$ model are expected to
be very fragile against quantum fluctuations, and indeed to leading order in $1/s$ the 
spin-wave correction to the spiral order parameter has been shown to diverge as $\log N$, 
where $N$ is the number of lattice sites \cite{Mulder:2010}, although it is still 
possible that higher-order terms in $1/s$ involving spin-wave interactions could stabilize
the spiral order for large enough values of $s$.  In view of the close proximity of
the classical collinear AFM aN state for small values of the NNNN coupling $J_3$ in the 
$J_{1}$--$J_{2}$--$J_{3}$ model, it seems very likely that spiral order in the
spin-$\frac{1}{2}$ $J_{1}$--$J_{2}$ model might well be totally absent.

To investigate this question further, and more generally to consider the entire
$T=0$ phase diagram of the $J_{1}$--$J_{2}$ model, we 
utilize the coupled cluster method (CCM) \cite{Bi:1991,Bi:1998,Fa:2004} 
as in our work for the analogous $J_{1}$--$J_{2}$--$J_{3}$ 
model with $J_{3}=J_{2}$ \cite{DJJF:2011_honeycomb}.
When used at high orders in the systematic approximation schemes developed for it, 
the CCM is an accurate approach to tackling a wide variety of quantum
spin systems \cite{j1j2_square_ccm1,Bi:2008_PRB,j1j2_square_ccm5,
Kr:2000,rachid05,Schm:2006,shastry3,richter10,
UJack_ccm,Reuther:2011_J1J2J3mod}. 
In particular, it can accurately locate the quantum critical points 
(QCPs) in such frustrated systems 
\cite{Bi:2008_PRB,j1j2_square_ccm5,rachid05,Schm:2006,
richter10,Reuther:2011_J1J2J3mod},
as well as helping to determine the nature of the phases involved, including
any quantum paramagnetic phases without magnetic order \cite{j1j2_square_ccm5}.

Since the CCM is a size-extensive method it provides results in the 
limit $N \rightarrow \infty$ from the outset.
However, it requires us to input a model (or reference) state, with
respect to which the quantum correlations may, in principle, be exactly 
included (and see, e.g., Refs. \cite{j1j2_square_ccm1,ccm2,ccm3} and 
references cited therein).  We use here the N\'{e}el (N), spiral, and 
anti-N\'{e}el (aN) states shown in Fig.~\ref{model} as our CCM
model states.  The CCM then incorporates multispin
correlations on top of the chosen gs model state $|\Phi\rangle$ for
the correlation operators $S$ and $\tilde{S}$ that parametrize the
exact gs ket and bra wave functions of the system in the respective
exponentiated forms $|\Psi \rangle=e^{S}|\Phi\rangle$ and 
$\langle \tilde{\Psi}|=\langle \Phi|\tilde{S}e^{-S}$, 
where $\langle \tilde{\Psi}|\Psi \rangle \equiv 1$.  The Schr\"{o}dinger 
ket and bra equations are $H|\Psi\rangle = E|\Psi\rangle$ 
and $\langle\tilde{\Psi}|H=E\langle\tilde{\Psi}|$ respectively.  
The correlation operators are defined as $S =
\sum_{I\neq0}{\cal S}_{I}C^{+}_{I}$ and $\tilde{S} = 1 +
\sum_{I\neq0}\tilde{\cal S}_{I}C^{-}_{I}$ respectively.  The 
operators $C^{+}_{I} \equiv (C^{-}_{I})^{\dagger}$ and  $C^{-}_{I}$ are the creation
and destruction operators respectively, where $C^{+}_{0} \equiv 1$ and
$\langle\Phi|C^{+}_{I} = 0\,;\, \forall I \neq 0$.  
The set $\{C^{+}_{I} \equiv s^{+}_{j_1}s^{+}_{j_2} \cdots s^{+}_{j_n}\}$, 
where $s^{+}_{j} \equiv s^{x}_{j} + is^{y}_{j}$, forms a
complete set of multispin creation operators with respect to the model
state $|\Phi\rangle$ as a generalized vacuum.  
We then calculate the correlation coefficients $\{{\cal S}_{I},\tilde{\cal S}_{I}\}$
by minimizing the gs energy expectation value $\bar{H} \equiv
\langle\tilde{\Psi}|H|\Psi\rangle$ with respect to
each of them.  This yields the coupled sets of equations $\langle
\Phi|C^{-}_{I}\mbox{e}^{-S}H\mbox{e}^{S}|\Phi\rangle = 0$ and
$\langle\Phi|\tilde{S}(\mbox{e}^{-S}H\mbox{e}^{S} -
E)C^{+}_{I}|\Phi\rangle = 0\,;\, \forall I \neq 0$, which are used to
calculate the ket- and bra-state correlation coefficients 
within specific truncation schemes on the retained set $\{I\}$ described below.  It is necessary to
use parallel computing routines for high-order computation \cite{ccm,ccm2,ccm3}.

For the $s=\frac{1}{2}$ case considered here we use the well-tested localized
LSUB$m$ truncation scheme which includes all multi-spin correlations
in the CCM correlation operators over all regions on the lattice
defined by $m$ or fewer contiguous lattice sites.  The
numbers $N_{f}$ of such fundamental configurations that are distinct under the symmetries of the
lattice and the model state in various LSUB$m$ approximations increases
rapidly with the truncation index $m$.  For example,
the highest LSUB$m$ level that we can reach, even with
massive parallelization and the use of supercomputing resources, is
LSUB$12$, for which $N_{f} = 293309$ for the aN state. 
The raw LSUB$m$ data still need to be extrapolated
to the exact $m \rightarrow \infty$ limit.  Although there are no exact extrapolation
rules we have a great deal of experience in doing so.
Thus, for the gs energy per spin, $E/N$, we use (see, e.g., Refs.,
\cite{rachid05,j1j2_square_ccm5,Schm:2006,Bi:2008_PRB,richter10,ccm3})
\begin{equation}
E(m)/N = a_{0}+a_{1}m^{-2}+a_{2}m^{-4}\,;     \label{E_extrapo}
\end{equation}
while for the magnetic order parameter (sublattice magnetization), $M$,
we use either the scheme
\begin{equation}
M(m) = b_{0}+b_{1}m^{-1}+b_{2}m^{-2}\,,    \label{M_extrapo_standard}
\end{equation}
for systems showing no or only slight frustration
(see, e.g., Refs.~\cite{rachid05,Kr:2000}), or the scheme
\begin{equation}
M(m) = c_{0}+c_{1}m^{-1/2}+b_{2}m^{-3/2}\,,   \label{M_extrapo_frustrated}
\end{equation}
for more strongly frustrated systems or ones showing a gs
order-disorder transition (see, e.g.,
Refs.~\cite{Bi:2008_PRB,j1j2_square_ccm5,richter10,Reuther:2011_J1J2J3mod}).
Since the hexagon is an important structural element of the honeycomb
lattice we perform the extrapolations only for LSUB$m$
data with $m \geq 6$.  

\begin{figure}[!tb]
\includegraphics[angle=270,width=8cm]{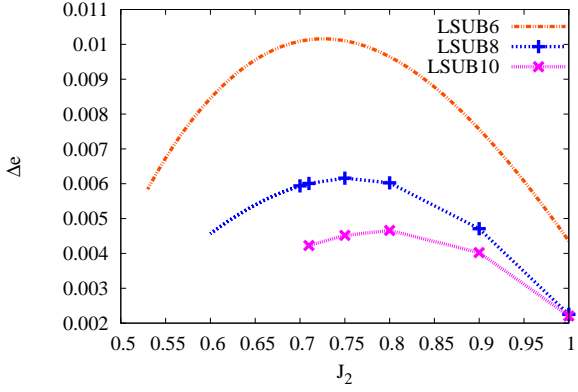}
\caption{Difference between the gs energies per spin ($e \equiv E/N$)
  of the spiral and anti-N\'{e}el phases ($\Delta e \equiv e^{{\rm
      spiral}}- e^{{\rm aN}}$) versus $J_{2}$ for the spin-$\frac{1}{2}$
  $J_{1}$--$J_{2}$ honeycomb model 
  ($J_{1}=1$) in LSUB$m$ approximations with
  $m=\{6,8,10\}$.}
\label{Ediff}
\end{figure}

We show in Fig.~\ref{Ediff} 
the difference in the gs energies of
the CCM LSUB$m$ results based on the spiral and aN model states.  For the
spiral state the pitch angle at a given LSUB$m$ level 
is chosen to minimize the corresponding estimate for the gs energy.
Although the energy differences are small the results for all values of $m$, as 
well as the extrapolated results, show clearly that, as expected from our previous
discussion, the spiral state that is the classical gs for all values $\kappa \equiv J_{2}/J_{1}>0.5$ 
gives way to the collinear aN state as the stable gs phase for the spin-$\frac{1}{2}$ model 
out to much higher values of $\kappa$.  If a quantum phase transition between
the spiral and aN states does exist, Fig.~\ref{Ediff} shows that it can occur 
only at a value $\kappa >1$.  Henceforth we concentrate on gs phases 
other than the spiral phase.

\begin{figure}[!bt]
\includegraphics[angle=270,width=8cm]{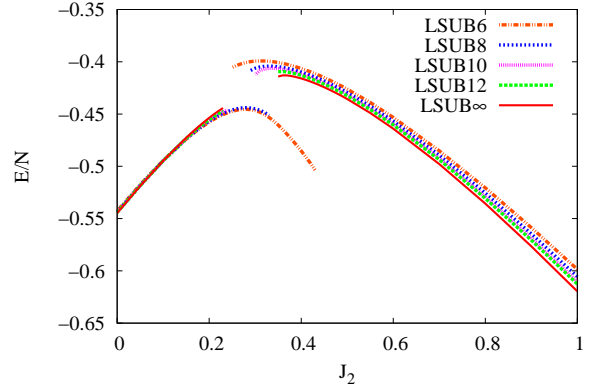}
\caption{CCM LSUB$m$ results for the gs energy, $E/N$, of the N\'{e}el
  and anti-N\'{e}el phases of the spin-$\frac{1}{2}$ $J_{1}$--$J_{2}$
  honeycomb model ($J_{1}=1$), with
  $m=\{6,8,10,12\}$ and the extrapolated LSUB$\infty$ result using this
  data set.}
\label{E}
\end{figure}

In Fig.~\ref{E}
we show results for the gs energy per spin, $E/N$, using
the N and aN states as model states.  We observe that each of
the energy curves based on a particular model state terminates at some
critical value of $\kappa$ (that itself depends on the LSUB$m$ 
approximation used), beyond which no real CCM solution can be found.  We
note that in Fig.~\ref{Ediff} results are shown for each LSUB$m$ case down 
to values of $\kappa$ at which real solutions based on the spiral 
model state cease to exist.  In all cases the corresponding termination point
at a given LSUB$m$ level shown in Fig.~\ref{E} for aN model
state is lower than that for the equivalent spiral model state case.
Such terminations of the CCM solutions are well understood \cite{Fa:2004,UJack_ccm}. 
They are simply manifestations of the quantum
phase transitions in the real system, and may thus be used to estimate the
positions of the corresponding QCPs \cite{Fa:2004}, 
although we do not do so here since we have more accurate criteria 
available as discussed below.  We note, however, that as is usually the case, 
the CCM LSUB$m$ results for finite $m$ values for
both the N and aN phases shown in Fig.~\ref{E} extend beyond the corresponding
LSUB$\infty$ transition points.  For large values of $m$ each LSUB$m$
transition point is quite close to the actual QCP 
where that phase ends.  For example, the LSUB12 termination points
shown in Fig.~\ref{E} are at $\kappa ^{\rm N}_{t} \approx 0.23$ for the N state
and $\kappa ^{\rm aN}_{t} \approx 0.35$ for the aN state.  The CCM
results show a clear intermediate regime in which neither of the 
quasiclassical AFM states (N or aN) is stable.
 
\begin{figure}[!tb]
  \includegraphics[angle=270,width=8cm]{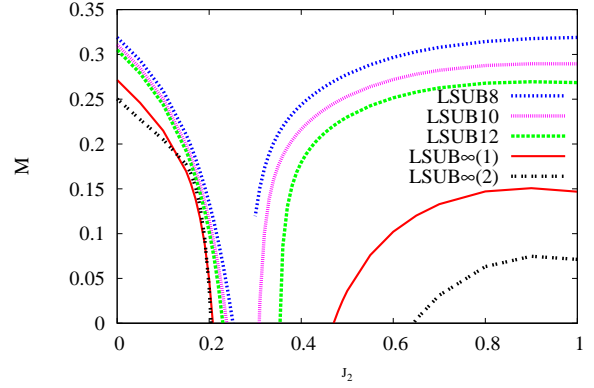}
  \caption{CCM LSUB$m$ results for the gs order parameter, $M$, 
   of the N\'{e}el and anti-N\'{e}el phases  
   of the spin-$\frac{1}{2}$ $J_{1}$--$J_{2}$
   honeycomb model ($J_{1}=1$), 
   with $m=\{8,10,12\}$, and the extrapolated LSUB$\infty$(1) 
   and LSUB$\infty$(2) results using this
   data set and Eqs. (\ref{M_extrapo_standard}) and 
   (\ref{M_extrapo_frustrated}) respectively.}  
\label{M}
\end{figure}

We now discuss the magnetic order parameter, $M$, in order to
investigate the stability of quasiclassical magnetic LRO.  
Our CCM results for $M$ are shown in Fig.~\ref{M}.
The extrapolated N\'{e}el order parameter goes to zero at a value
$\kappa _{c_1}$ that is very insensitive both to which extrapolation scheme
is used of Eqs. (\ref{M_extrapo_standard}) or (\ref{M_extrapo_frustrated}), and
to whether or not we include the LSUB6 data point in the extrapolations.  Our
best estimate is $\kappa _{c_1} \approx 0.207 \pm 0.003$.  This value can be
considered as our first CCM estimate of the corresponding QCP of the model.  It 
is in reasonable agreement with, but much more precise than, similar estimates
for $\kappa _{c_1}$ of about 0.17-0.22 from an exact diagonalization (ED) approach \cite{Albuquerque:2011},
about 0.2 from an alternative ED approach \cite{honey5}, about 0.2 from a Schwinger 
boson approach \cite{honey1}, and about 0.13-0.17 from a pseudo-fermion functional
renormalization group approach \cite{honey7}, but in substantial disagreement with
a recent variational Monte Carlo (VMC) estimate of about 0.08 \cite{Clark:2011_honeyVMC}.
As the authors admit, the VMC study seems to substantially underestimate the QCP 
$\kappa _{c_1}$ at which N\'{e}el order disappears.  As expected, our
own estimate shows that, as usual, quantum fluctuations preserve the collinear
N\'{e}el order to stronger frustrations than the corresponding classical 
transition to noncollinear spiral order at $\kappa _{\rm cl}=\frac{1}{6}$.
  
By contrast with the situation at the lower QCP at $\kappa _{c_1}$, 
Fig.~\ref{M} shows that the corresponding QCP at $\kappa _{c_2}$ at
which the anti-N\'{e}el order vanishes is considerably more difficult to
estimate from the extrapolated CCM LSUB$m$ values, with estimates that
range from 0.47 to 0.64.  We find a much more accurate estimate for 
$\kappa _{c_2}$ below.  Nevertheless it is clear already that a new quantum phase
exists in the range $\kappa _{c_1}<\kappa<\kappa _{c_2}$. It is also clear 
that, as suggested above, the two QCPs are very close to the corresponding 
CCM termination points seen in Fig.~\ref{E}.  In our previous work
on the spin-$\frac{1}{2}$ $J_{1}$--$J_{2}$--$J_{3}$ model on the honeycomb lattice 
\cite{DJJF:2011_honeycomb} with $J_{1}=1$, we found strong evidence
for a nonmagnetic plaquette valence-bond crystal (PVBC) phase along the line
$J_{3}=J_{2}$ bewteen the two quasiclassical AFM phases, namely the N\'{e}el (N)
and striped (S) phases.  It seems likely that the corresponding phase in the 
present $J_{3}=0$ case, which now intervenes between the N and aN phases, might
also be the same PVBC phase.

\begin{figure}[!t]
\begin{center}
\mbox{
\subfloat{\includegraphics[width=6cm,height=6cm,angle=270]{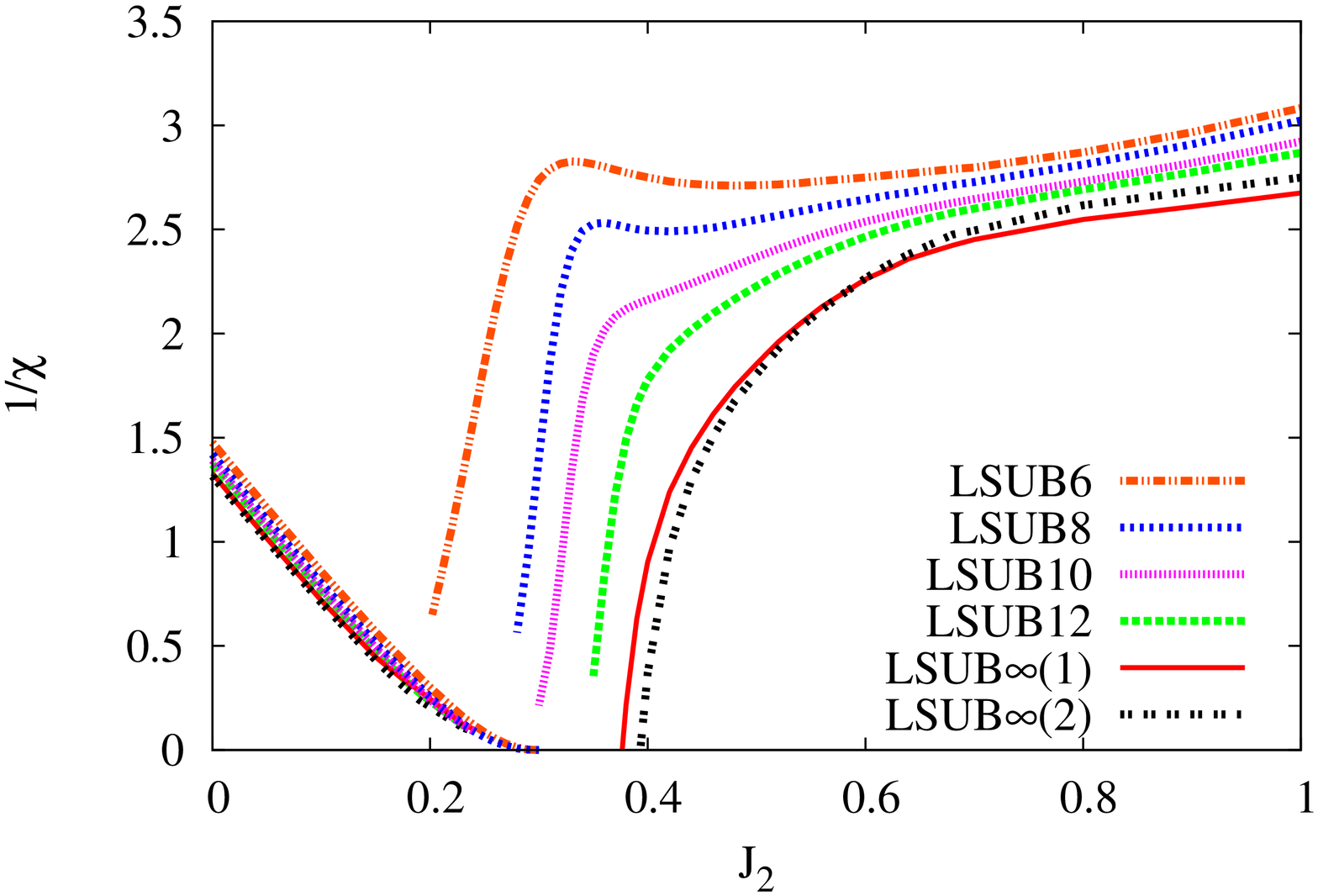}}
\raisebox{-3.5cm}{
\subfloat{\includegraphics[width=2.2cm,height=2.2cm]{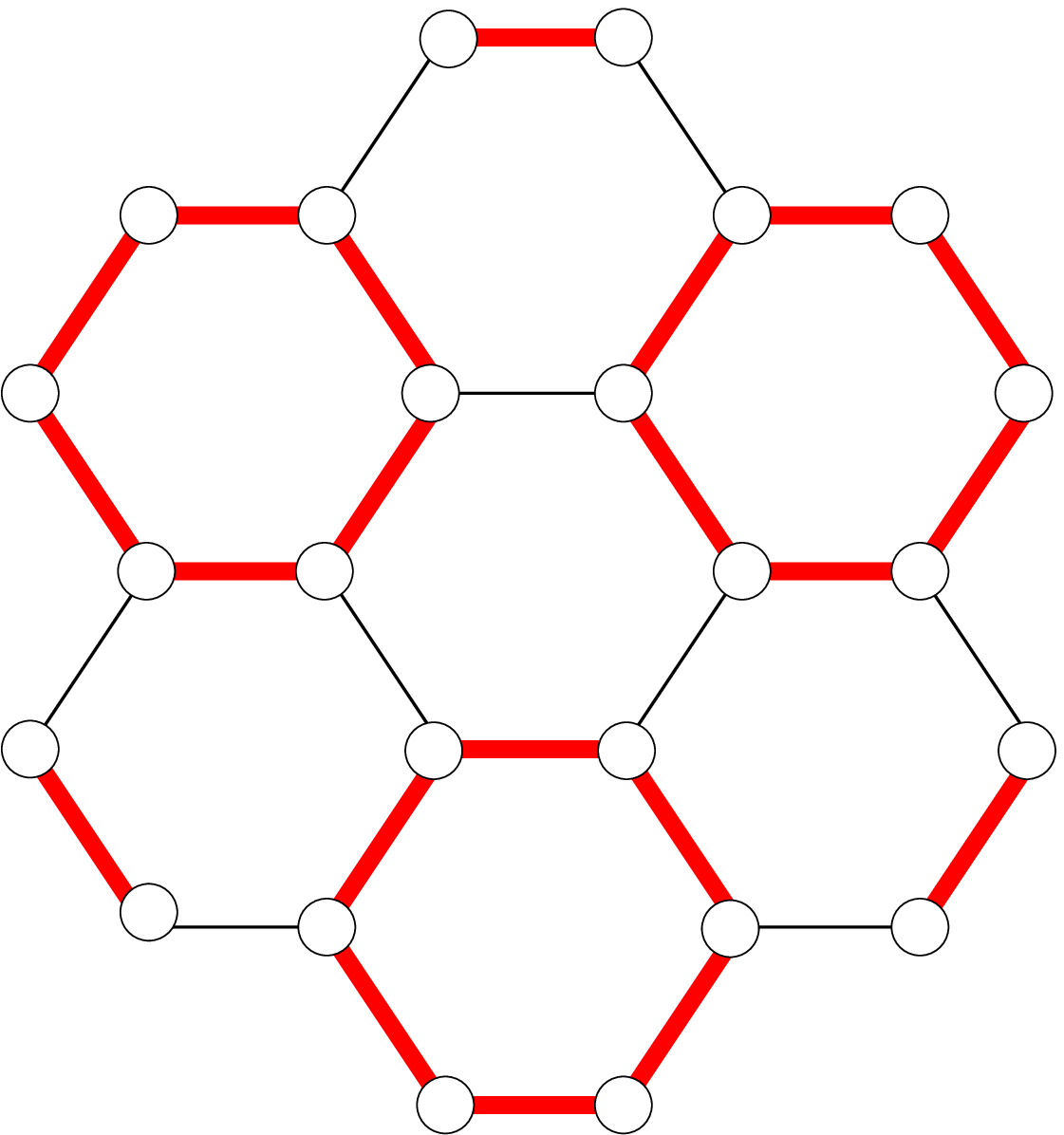}}
}
}
\caption{(Color online) Left: CCM LSUB$m$ results for the inverse plaquette
  susceptibility, $1/\chi$, of the N\'{e}el
  and anti-N\'{e}el phases of the spin-$\frac{1}{2}$ $J_{1}$--$J_{2}$
  honeycomb model ($J_{1}=1$) with $m=\{6,8,10,12\}$, and the extrapolated results
  LSUB$\infty$(1) using $m=\{6,8,10,12\}$ and LSUB$\infty$(2) using
  $m=\{8,10,12\}$ (see text).  Right: The 
  fields $F=\delta\; \hat{O}$ for the plaquette susceptibility
  $\chi$.  Thick (red) and thin (black) lines correspond respectively
  to strengthened and weakened NN exchange couplings, where $\hat{O} =
  \sum_{\langle i,j \rangle} a_{ij}
  \mathbf{s}_{i}\cdot\mathbf{s}_{j}$, and the sum runs over all NN
  bonds, with $a_{ij}=+1$ and $-1$ for thick (red) and thin (black)
  lines respectively.}
\label{X}\
\end{center}
\end{figure}

In order to investigate the possibility of a PBVC phase we consider a
generalized susceptibility $\chi_F$ that describes the response of the
system to a ``field'' operator $F$ (see, e.g.,
Ref.~\cite{j1j2_square_ccm5}).  A field term $F=\delta\; \hat{O}$
is added to the Hamiltonian (\ref{eq1}), where $\hat{O}$ is an
operator which corresponds here to the possible PVBC order 
illustrated in Fig.~\ref{X}, and which thus
breaks the translational symmetry of $H$.  The energy per site
$E(\delta)/N=e(\delta)$ is then calculated in the CCM for the perturbed
Hamiltonian $H+F$, using both the N and aN model states.
The susceptibility is defined as $\chi_{F} \equiv -
\left. (\partial^2{e(\delta)})/(\partial {\delta}^2)
\right|_{\delta=0}$.  Clearly,  the gs phase becomes 
unstable against the perturbation $F$ when
$\chi_F^{-1}$ becomes zero. As in Ref. \cite{DJJF:2011_honeycomb} we
use the extrapolation scheme 
$\chi_{F}^{-1}(m) = d_{0}+d_{1}m^{-2}+d_{2}m^{-4}$.  

Our CCM results for $\chi_{F}^{-1}$ are shown in Fig.~\ref{X}.  The
number of LSUB12 fundamental configurations for the plaquette susceptibility 
for the aN state is $N_{f} = 877315$.  The extrapolated inverse
susceptibility vanishes on the N\'eel side at $\kappa \approx 0.24 \pm
0.01$ and on the anti-N\'{e}el side at $\kappa \approx 0.385 \pm
0.010$.  The shape of the CCM curves where $\chi_F^{-1} \rightarrow 0$
on the N\'{e}el side is strongly suggestive of a continuous transition
there, just as we found for the corresponding
$J_{1}$--$J_{2}$--$J_{3}$ model \cite{DJJF:2011_honeycomb}.  The
shallow slope of the $\chi_F^{-1}$ curves there makes it difficult to
estimate accurately the point where it vanishes.  Nevertheless it is
certainly consistent with the much more accurate value we obtained for
$\kappa _{c_1}$ above from the point where the N\'{e}el order
parameter $M$ vanishes. On the other hand we cannot exclude the
possibility of the transition between the N\'{e}el and PVBC states
occurring via an intermediate phase that exists in the region
$\kappa_{c_1} < \kappa \lesssim 0.24$.  Just such an intermediate
(resonating valence bond spin-liquid) state has been discussed in Ref.
\cite{honey2}.  By contrast, the shape of the CCM curves for
$\chi_F^{-1}$ on the anti-N\'{e}el side are much more indicative of a
first-order transition, and the point where $\chi_F^{-1} \rightarrow
0$ on that side gives us our best estimate for $\kappa_{c_2} \approx
0.385 \pm 0.010$.  Again, this value is in good agreement with, but
much more accurate than, estimates of about 0.35-0.4 from two
different ED studies \cite{honey5,Albuquerque:2011}.  On both the N
and aN sides it seems very likely that the PVBC phase occurs at, or
very close to, the QCPs where the quasiclassical magnetic LRO in the N
and aN phases vanishes.  Since the N and PVBC phases break different
symmetries, and our CCM results show that they appear to meet at
$\kappa_{c_1} \approx 0.21$ at a continuous transition, they support
the deconfinement scenario there.  The possibility of deconfined
quantum criticality for the frustrated honeycomb HAFM was pointed out
in Ref. \cite{Senthil:2004}, and supporting evidence for the
$J_{1}$--$J_{2}$--$J_{3}$ model has also been found both by us
\cite{DJJF:2011_honeycomb} and by others \cite{Albuquerque:2011}.

We have studied the influence of quantum spin
fluctuations on the gs properties of the spin-$\frac{1}{2}$
$J_{1}$--$J_{2}$ HAFM ($J_{1}>0, \kappa \equiv J_{2}/J_{1}$) 
on the honeycomb lattice.  We find a paramagnetic
PVBC phase in the regime $\kappa_{c_1}<\kappa<\kappa_{c_2}$, where
$\kappa_{c_1} \approx 0.207 \pm 0.003$ and $\kappa_{c_2} \approx 0.385 \pm 0.010$.
The transition at $\kappa_{c_1}$ to the N\'{e}el phase appears to
be a continuous transition of the deconfinement variety, while that at
$\kappa_{c_2}$ is of first-order type to an anti-N\'{e}el-ordered AFM
phase that does not occur in the classical version of the model
except at the critical point $\kappa = \frac{1}{2}$.  Our results indicate that the spiral
phases that exist classically for all values $\kappa > \frac{1}{6}$ are absent
for all values $\kappa \lesssim 1$, but may exist for larger values.  To
investigate this and other aspects of the model further we shall
present results in a future paper on the phase diagram of the
extended $J_{1}$--$J_{2}$--$J_{3}$ model, using the same CCM techniques.

We thank the University of Minnesota Supercomputing Institute for the grant of
supercomputing facilities.


\begin{thebibliography}{200}

\bibitem{Senthil:2004}
\Name{Senthil T., Vishwanath A., Balents L., Sachdev S. \and Fisher M. P. A.}
\Review{Science}
\Vol{303}
\Year{2004}
\Page{1490}.

\bibitem{Moessner:2006}
\Name{Moessner R. \and Ramirez A. P.}
\Review{Phys. Today}
\Vol{59}
\Year{February 2006}
\Page{24}.

\bibitem{SRFB:2004}
\Book{Quantum Magnetism}
\Editor{Schollw{\"{o}}ck U., Richter J., Farnell D. J. J. \and Bishop R. F.},
{\em Lecture Notes in Physics} 
\Vol{645}
\Publ{Springer-Verlag, Berlin} 
\Year{2004}.

\bibitem{Sachdev_book:1999}
\Book{Quantum Phase Transitions,}
\Name{Sachdev S.}
\Publ{Cambridge Univ. Press, Cambridge}
\Year{1999}.

\bibitem{j1j2_square_ccm1} 
\Name{Bishop R. F., Farnell D. J. J. \and Parkinson J. B.} 
\Review{Phys. Rev. B} 
\Vol{58} 
\Year{1998} 
\Page{6394}.

\bibitem{Bi:2008_PRB} 
\Name{Bishop R. F., Li P. H. Y., Darradi R., Schulenburg J. \and Richter J.}
\Review{Phys.\ Rev.\ B} 
\Vol{78} 
\Year{2008} 
\Page{054412}.

\bibitem{j1j2_square_ccm5}
\Name{Darradi R., Derzhko O., Zinke R., Schulenburg J., Kr\"uger S. E. \and Richter J.}
\Review{Phys.~Rev.~B} 
\Vol{78} 
\Year{2008} 
\Page{214415}.

\bibitem{Bernu:1994}
\Name{Bernu B., Lhuillier C. \and Pierre L.}
\Review{Phys.~Rev.~Lett.} 
\Vol{69} 
\Year{1992} 
\Page{2590}.

\bibitem{Kagome_Schn:2008}
\Name{Schnyder A. P., Starykh O. A. \and L. Balents}
\Review{Phys. Rev. B} 
\Vol{78} 
\Year{2008} 
\Page{174420}.

\bibitem{honey1}  
\Name{Mattsson A., Fr\"ojdh P. \and Einarsson T.}
\Review{Phys. Rev. B} 
\Vol{49} 
\Year{1994} 
\Page{3997}. 

\bibitem{honey2}
\Name{Fouet J. B., Sindzingre P. \and Lhuillier C.} 
\Review{Eur. Phys. J. B} 
\Vol{20} 
\Year{2001} 
\Page{241}. 

\bibitem{exp} 
\Name{Okubo S., Elmasry F., Zhang W., Fujisawa M.,
Sakurai T., Ohta H., Azuma M., Sumirnova O. A. \and Kumada N.}
\Review{J. Phys.: Conf. Series} 
\Vol{200} 
\Year{2010} 
\Page{022042}.

\bibitem{honey5}
\Name{Mosadeq H., Shahbazi F. \and Jafari S. A.}
\Review{J. Phys.: Condens. Matter} 
\Vol{23} 
\Year{2011} 
\Page{226006}.

\bibitem{Cabra}
\Name{Cabra D. C., Lamas C. A. \and Rosales H. D.}
\Review{Phys. Rev. B} 
\Vol{83} 
\Year{2011} 
\Page{094506}.

\bibitem{honey7}
\Name{Reuther J., Abanin D. A. \and Thomale R.} 
\Review{Phys. Rev. B} 
\Vol{84}
\Year{2011}
\Page{014417}.

\bibitem{Albuquerque:2011}
\Name{Albuquerque A. F., Schwandt D., Het\'{e}nyi B., Capponi S., Mambrini M. \and L\"auchli A. M.}
\Review{Phys. Rev. B} 
\Vol{84} 
\Year{2011} 
\Page{024406}.   

\bibitem{kitaev} 
\Name{Kitaev A.}
\Review{Ann. Phys. (N.Y.)} 
\Vol{321} 
\Year{2006} 
\Page{2}.

\bibitem{meng}
\Name{Meng Z. Y., Lang T. C., Wessel S., Assaad F. F. \and Muramatsu A.} 
\Review{Nature} 
\Vol{464} 
\Year{2010} 
\Page{847}.

\bibitem{Miura:2006}
\Name{Miura Y., Hiari R., Kobayashi Y. \and Sato M.}
\Review{J. Phys. Soc. Jpn.} 
\Vol{75} 
\Year{2006} 
\Page{084707}.

\bibitem{Kataev:2005}
\Name{Kataev V., M\"oller A., L\"ow U., Jung W., Schittner N., Kriener M. \and Freimuth A.}
\Review{J. Magn. Magn. Mater.} 
\Vol{290-291} 
\Year{2005} 
\Page{310}.

\bibitem{Regnault}
\Name{Regnault L. P. \and Rossat-Mignod J.}
in \Book{Phase Transitions in Quasi-Two-Dimensional Planar Magnets}
\Editor{De Jongh L. J.}
\Publ{Kluwer Academic Publshers} 
\Year{1990} 
\Page{271}.

\bibitem{Tsirlin:2010}
\Name{Tsirlin A. A., Janson O. \and Rosner H.}
\Review{Phys. Rev. B} 
\Vol{82} 
\Year{2010} 
\Page{144416}.

\bibitem{DJJF:2011_honeycomb}
\Name{Farnell D. J. J., Bishop R. F., Li P. H. Y., Richter J. \and Campbell C. E.}
\Review{Phys. Rev. B} 
\Vol{84} 
\Year{2011} 
\Page{012403}.

\bibitem{PL:2011_Honeycomb_FM}
\Name{Li P. H. Y., Bishop R. F., Farnell D. J. J., Richter J. \and Campbell C. E.}
\Review{preprint arXiv:1109.6229} 
\Year{2011}.

\bibitem{Rastelli:1979}
\Name{Rastelli E., Tassi A. \and Reatto L.}
\Review{Physica} 
\Vol{97B}
\Year{1979}
\Page{1}.

\bibitem{Mulder:2010}
\Name{Mulder A., Ganesh R., Capriotti L. \and Paramekanti A.}
\Review{Phys. Rev. B} 
\Vol{81} 
\Year{2010} 
\Page{214419}.

\bibitem{Bi:1991}
\Name{Bishop R. F.} 
\Review{Theor. Chim. Acta} 
\Vol{80} 
\Year{1991} 
\Page{95}.

\bibitem{Bi:1998} 
\Name{Bishop R. F.} in
\Book{Microscopic Quantum Many-Body Theories and Their Applications}
\Editor{Navarro J. \and Polls A.} 
\Vol{510} of {\em Lecture Notes in Physics} 
\Publ{Springer-Verlag, Berlin}
\Year{1998}, p.1.

\bibitem{Fa:2004} 
\Name{Farnell D. J. J. \and Bishop R. F.} in 
\Book{Quantum Magnetism} \Editor{Schollw{\"{o}}ck U., Richter J., Farnell D. J. J. \and Bishop R. F.}
\Vol{645} of {\em Lecture Notes in Physics} 
\Publ{Springer-Verlag, Berlin} 
\Year{2004}, p.307.

\bibitem{Kr:2000} 
\Name{Kr{\"{u}}ger S. E., Richter J., Schulenburg J.,
Farnell D. J. J. \and Bishop R. F.} 
\Review{Phys. Rev. B} 
\Vol{61} 
\Year{2000} 
\Page{14607}. 

\bibitem{rachid05} 
\Name{Darradi R., Richter J. \and Farnell D. J. J.}
\Review{Phys. Rev. B} 
\Vol{72} 
\Year{2005} 
\Page{104425}.

\bibitem{Schm:2006} 
\Name{Schmalfu\ss D., Darradi R., Richter J., Schulenburg J. \and Ihle D.}
\Review{Phys. Rev. Lett.} 
\Vol{97} 
\Year{2006} 
\Page{157201}.

\bibitem{shastry3} 
\Name{Farnell D. J. J., Richter J., Zinke R. \and Bishop R.~F.}
\Review{J. Stat. Phys.} 
\Vol{135} 
\Year{2009} 
\Page{175}.

\bibitem{richter10}  
\Name{Richter J., Darradi R., Schulenburg J., Farnell D. J. J. \and Rosner H.}
\Review{Phys. Rev. B} 
\Vol{81} 
\Year{2010} 
\Page{174429}.

\bibitem{UJack_ccm}
\Name{Bishop R. F., Li P. H. Y., Farnell D. J. J. \and Campbell C. E.}
\Review{Phys. Rev. B} 
\Vol{82} 
\Year{2010} 
\Page{024416}.

\bibitem{Reuther:2011_J1J2J3mod}
\Name{Reuther J., W\"{o}lfle P., Darradi R., Brenig W., Arlego M. \and Richter J.}
\Review{Phys. Rev. B} 
\Vol{83} 
\Year{2011} 
\Page{064416}.

\bibitem{ccm2} 
\Name{Zeng C., Farnell D. J. J. \and Bishop R. F.}
\Review{J. Stat. Phys.} 
\Vol{90} 
\Year{1998} 
\Page{327}.

\bibitem{ccm3} 
\Name{Bishop R. F., Farnell D. J. J., Kr\"uger S.E., Parkinson J. B., Richter J. \and Zeng C.} 
\Review{J. Phys.: Condens. Matter} 
\Vol{12} 
\Year{2000} 
\Page{6887}.

\bibitem{ccm} We use the program
package CCCM of D.~J.~J. Farnell and J. Schulenburg, see
http://www-e.uni-magdeburg.de/jschulen/ccm/index.html.

\bibitem{Clark:2011_honeyVMC}
\Name{Clark B. K., Abanin D. A. \and Sondhi S. L.}
\Review{Phys.~Rev.~Lett.} 
\Vol{107} 
\Year{2011} 
\Page{087204}.


\end{thebibliography}
\end{document}